\documentclass{emulateapj}
\usepackage{lscape}

\def\blender{{\tt BLENDER}}
\def\kepler{\emph{Kepler}}
\def\kms{\ifmmode{\rm km\thinspace s^{-1}}\else km\thinspace s$^{-1}$\fi}
\def\ms{\ifmmode{\rm m\thinspace s^{-1}}\else m\thinspace s$^{-1}$\fi}

\newcommand{\wspitzer}{\emph{Warm Spitzer}}
\newcommand{\spitzer}{\emph{Spitzer}}
\newcommand{\koi}{KOI-072.02}

\shortauthors{Fressin et al.}
\shorttitle{Kepler-10\,c}

\begin{document}


\title{Kepler-10\,c, a 2.2-Earth radius transiting planet in a
multiple system}

\author{
Fran\c{c}ois Fressin\altaffilmark{1},
Guillermo Torres\altaffilmark{1},
Jean-Michel D\'{e}sert\altaffilmark{1},
David Charbonneau\altaffilmark{1},
Natalie M.\ Batalha\altaffilmark{2},
Jonathan J.\ Fortney\altaffilmark{3},
Jason F.\ Rowe\altaffilmark{4,5},
Christopher Allen\altaffilmark{4,6},
William J.\ Borucki\altaffilmark{4},
Timothy M.\ Brown\altaffilmark{7},
Stephen T.\ Bryson\altaffilmark{4},
David R.\ Ciardi\altaffilmark{8},
William D.\ Cochran\altaffilmark{9},
Drake Deming\altaffilmark{10},
Edward W.\ Dunham\altaffilmark{11},
Daniel C.\ Fabrycky\altaffilmark{3},
Thomas N.\ Gautier III\altaffilmark{12},
Ronald L.\ Gilliland\altaffilmark{13},
Christopher E.\ Henze\altaffilmark{4},
Matthew J.\ Holman\altaffilmark{1},
Steve B.\ Howell\altaffilmark{14},
Jon M.\ Jenkins\altaffilmark{4,15},
Kamal Kamal\altaffilmark{4,6},
Karen Kinemuchi\altaffilmark{4,16},
Heather Knutson\altaffilmark{17},
David G.\ Koch\altaffilmark{8},
David W.\ Latham\altaffilmark{1},
Jack J.\ Lissauer\altaffilmark{4},
Geoffrey W.\ Marcy\altaffilmark{17},
Darin Ragozzine\altaffilmark{1},
Dimitar D.\ Sasselov\altaffilmark{1},
Martin Still\altaffilmark{4,16}, and
Peter Tenenbaum\altaffilmark{4,6}
}

\altaffiltext{1}{Harvard-Smithsonian Center for Astrophysics,
Cambridge, MA 02138, USA, e-mail: ffressin@cfa.harvard.edu}
\altaffiltext{2}{San Jose State University, San Jose, CA 95192, USA}
\altaffiltext{3}{University of California, Santa Cruz, CA 95064, USA}
\altaffiltext{4}{NASA Ames Research Center, Moffett Field, CA 94035, USA}
\altaffiltext{5}{NASA Postdoctoral Program Fellow}
\altaffiltext{6}{Uddin Orbital Sciences Corporation, Moffett Field, CA 94035, USA}
\altaffiltext{7}{Las Cumbres Observatory Global Telescope, Goleta, CA 93117, USA}
\altaffiltext{8}{NASA Exoplanet Science Institute/Caltech, Pasadena, CA USA 91125, USA}
\altaffiltext{9}{McDonald Observatory, University of Texas at Austin, Austin, TX 78712, USA}
\altaffiltext{10}{NASA Goddard Space Flight Center, Greenbelt MD 20771, USA}
\altaffiltext{11}{Lowell Observatory, Flagstaff, AZ 86001, USA}
\altaffiltext{12}{Jet Propulsion Laboratory/California Institute of Technology, Pasadena, CA 91109, USA}
\altaffiltext{13}{Space Telescope Science Institute, Baltimore, MD 21218, USA}
\altaffiltext{14}{National Optical Astronomy Observatory, Tuczon, AZ 85719, USA}
\altaffiltext{15}{SETI Institute/NASA Ames Research Center, Moffett Field, CA 94035, USA}
\altaffiltext{16}{Bay Area Environmental Research Institute, 560 Third St.\ West, Sonoma, CA 95476, USA}
\altaffiltext{17}{University of California, Berkeley, CA 94720, USA}

\begin{abstract}

The \kepler\ Mission has recently announced the discovery of
Kepler-10\,b, the smallest exoplanet discovered to date and the first
rocky planet found by the spacecraft.  A second, 45-day period
transit-like signal present in the photometry from the first eight
months of data could not be confirmed as being caused by a planet at
the time of that announcement. Here we apply the light-curve modeling
technique known as \blender\ to explore the possibility that the
signal might be due to an astrophysical false positive (blend).  To
aid in this analysis we report the observation of two transits with
the \spitzer\ Space Telescope at 4.5\,\micron. When combined they
yield a transit depth of $344 \pm 85$ ppm that is consistent with the
depth in the \kepler\ passband ($376 \pm 9$ ppm, ignoring limb
darkening), which rules out blends with an eclipsing binary of a
significantly different color than the target. Using these
observations along with other constraints from high-resolution imaging
and spectroscopy we are able to exclude the vast majority of possible
false positives. We assess the likelihood of the remaining blends, and
arrive conservatively at a false alarm rate of $1.6 \times 10^{-5}$
that is small enough to validate the candidate as a planet (designated
Kepler-10\,c) with a very high level of confidence. The radius of this
object is measured to be $R_p = 2.227_{-0.057}^{+0.052}\,R_{\earth}$
(in which the error includes the uncertainty in the stellar
properties), but currently available radial-velocity measurements only
place an upper limit on its mass of about 20\,$M_{\earth}$.
Kepler-10\,c represents another example (with Kepler-9\,d and
Kepler-11\,g) of statistical ``validation'' of a transiting exoplanet,
as opposed to the usual ``confirmation'' that can take place when the
Doppler signal is detected or transit timing variations are
measured. It is anticipated that many of \kepler's smaller candidates
will receive a similar treatment since dynamical confirmation may be
difficult or impractical with the sensitivity of current
instrumentation.


\end{abstract}


\section{Introduction}
\label{sec:introduction}

The \kepler\ Mission has recently made public a catalog of all
transiting planet candidates identified during the first four months
of observation by the spacecraft \citep{Borucki:11b}. Included in this
list of 1235 objects are nearly 300 in the category of super-Earths
(defined here as having radii in the range $1.25\,R_{\earth} < R_p <
2\,R_{\earth}$), and several dozen of Earth size ($R_p <
1.25\,R_{\earth}$).  The wealth of new information promises to
revolutionize our knowledge of extrasolar planets. Although strictly
speaking these are still only \emph{candidates} since confirmation by
spectroscopic or other means is not yet in hand, expectations are high
that the rate of false positives in this list is relatively small
\citep[see][]{Borucki:11b, Morton:11}.  Consequently, results from
this sample concerning the general properties of exoplanets have
already begun to emerge, including studies of the architecture and
dynamics of multiple transiting systems \citep{Lissauer:11b}, an
investigation of the statistical distribution of eccentricities
\citep{Moorhead:11}, and first estimates of the rate of occurrence of
planets larger than 2\,$R_{\earth}$ with orbital periods up to 50 days
\citep{Howard:11}, among others.

For good reasons the confirmation or ``validation'' of small
transiting planets\footnote{In the context of this paper
``confirmation'' refers to the unambiguous detection of the
gravitational influence of the planet on its host star or on other
bodies in the system (e.g., the Doppler signal, or transit timing
variations) to establish the planetary nature of the candidate; when
this is not possible, we speak of ``validation'', which involves an
estimate of the false alarm probability.}  (Earth-size or
super-Earth-size) has attracted considerable attention, but has proven
to be non-trivial in many cases because of the difficulty of detecting
the tiny radial-velocity (RV) signatures that these objects cause on
their parent stars, as exemplified by the cases of CoRoT-7\,b
\citep{Leger:09}, Kepler-9\,d \citep{Torres:11}, and Kepler-11\,g
\citep{Lissauer:11a}.  In fact, such spectroscopic signals are often
too small to detect with current instrumentation, and the planetary
nature of the candidate must be established statistically, as in the
latter two cases.

The smallest planet discovered to date, Kepler-10\,b, was announced
recently by \cite{Batalha:11}, and is the \kepler\ Mission's first
rocky planet. It has a measured radius of $1.416^{+0.033}_{-0.036}
\,R_{\earth}$ and a mass of $4.6^{+1.2}_{-1.3}\,M_{\earth}$, leading
to a mean density of $8.8^{+2.1}_{-2.9}\,{\rm g~cm}^{-3}$ that implies
a significant iron mass fraction \citep{Batalha:11}.  Its parent star,
Kepler-10 (KIC\,11904151, 2MASS\,119024305+5014286), is relatively
bright among the \kepler\ targets (\kepler\ magnitude $K\!p = 10.96$)
and displays \emph{two} periodic signals with periods of 0.84 days and
45.3 days, and flux decrements (ignoring limb darkening) of $152 \pm
4$ ppm and $376 \pm 9$ ppm, respectively \citep{Batalha:11}. The
extensive observations that followed the detection of these signals
are documented in detail by those authors, and include the difficult
measurement of the reflex radial-velocity motion of the star with a
semi-amplitude of only $3.3^{+0.8}_{-1.0}$\,\ms\ and a period that is
consistent with the shorter signal.  As is customary also in
ground-based searches for transiting planets, the shapes of the
spectral lines were examined carefully to rule out changes of similar
amplitude correlating with orbital phase that might indicate a false
positive, such as a background eclipsing binary (EB) blended with the
target, or an EB physically associated with it.  However the precision
of the measurements (bisector spans) compared to the small RV
amplitude did not allow such changes to be ruled out unambiguously.
False positive scenarios were explored with the aid of \blender, a
technique that models the transit light curves to test a wide range of
blend configurations \citep{Torres:11}, and it was found that the
overwhelming majority of them can be rejected.  This and other
evidence presented by \cite{Batalha:11} allowed the planetary nature
of Kepler-10\,b to be established with very high confidence.

This was not the case, however, for the 45-day period signal referred
to as \koi\ (\kepler\ Object of Interest 72.02), which is the subject
of this paper. No significant RV signal was detected at this period,
and only an upper limit on its amplitude could be placed.  Using
\blender, \cite{Batalha:11} were able to rule out a large fraction of
the blend scenarios involving circular orbits (including hierarchical
triples), but eccentric orbits were not explored because of the
increased complexity of the problem and the much larger space of
parameters for false positives.  While circular orbits are a
reasonable assumption for Kepler-10\,b because of the strong effects
of tidal forces at close range, this is not true for \koi\ on account
of its much longer orbital period \citep[see, e.g.,][]{Mazeh:08};
eccentric orbits can not be ruled out.

This provides the motivation for the present work, in which we set out
to examine all viable astrophysical false positive scenarios for \koi\
with the goal of validating it as a bona-fide planet. In addition to
improvements in the \blender\ modeling, we bring to bear new
near-infrared observations obtained with the \spitzer\ Space Telescope
in which the transits are clearly detected, as well as the complete
arsenal of follow-up observations gathered by the \kepler\ team,
including high-resolution adaptive optics imaging and speckle
interferometry, high-resolution spectroscopy, and an analysis based on
the \kepler\ observations themselves of the difference images in and
out of transit for positional displacements (centroid motion). All of
these observations combined with the strong constraints provided by
\blender\ significantly limit the kinds of blends that remain
possible, and as we describe below they allow us to claim with very
high confidence that \koi\ is indeed a planet. Its estimated radius is
approximately 60\% of that of Neptune. With this, Kepler-10 becomes
the Mission's third confirmed multi-planet system \citep[after
Kepler-9 and Kepler-11;][]{Holman:10, Lissauer:11a} containing a
transiting super-Earth-size planet and at least one larger planet that
also transits.

We begin with a brief recapitulation of the \blender\ technique,
including recent improvements. We then present the \wspitzer\
observations at 4.5\,\micron\ that help rule out many blends, and we
summarize additional constraints available from other observations.
This is followed by the application of \blender\ to \koi\ in order to
identify all blends scenarios that can mimic the \kepler\ transit
light curve.  Next we combine this information with the other
constraints and carry out a statistical assessment of the false alarm
rate for the planet hypothesis, leading to the validation of the
candidate as Kepler-10\,c.  We conclude with a discussion of the
possible constitution of the new planet in the light of current
models, and the significance of this type of validation.

\section{Rejecting false positives with \blender}
\label{sec:blender}

The detailed morphology of a transit light curve (length of
ingress/egress, total duration) contains important information that
can be used to reject many false positive scenarios producing
brightness variations that do not quite have the right shape, even
though they may well match the observed transit depth \citep[see,
e.g.,][]{Snellen:09}.  \blender\ \citep{Torres:04, Torres:11} takes
advantage of this to explore a very large range of scenarios,
including background or foreground eclipsing binaries blended with the
target, as well as eclipsing binaries physically associated with the
target in a hierarchical triple configuration. Following the notation
introduced by \cite{Torres:11}, the objects composing the binary are
referred to as the ``secondary'' and ``tertiary'', and the candidate
is the ``primary''. The tertiary can be either a star (including a
white dwarf) or a planet, and the secondary can be a main-sequence
star or a (background) giant.

With the help of model isochrones to set the stellar properties,
\blender\ simulates blend light curves resulting from the flux of the
eclipsing pair diluted by the brighter target (and any additional
stars that may fall within the photometric aperture). Each simulated
light curve is compared with the \kepler\ observations in a $\chi^2$
sense to identify which of them result in acceptable fits (to be
defined later). The parameters varied during the simulations are the
mass of the secondary star ($M_2$), the mass of the tertiary ($M_3$,
or its radius $R_3$ if a planet), the impact parameter ($b$), the
relative linear distance ($d$) between the eclipsing pair and the
target, and the relative duration ($D/D_{\rm circ}$) of the transit
compared to the duration for a circular orbit (see below). For
convenience the relative linear distance is parametrized in terms of
the difference in distance modulus, $\Delta\delta$, where
$\Delta\delta = 5\log(d_{\rm EB}/d_{\rm KOI})$.  In the case of
hierarchical triple configurations the isochrone for the binary is
assumed to be the same as for the primary \citep[metallicity of ${\rm
[Fe/H]} = -0.15$ and a nominal age of 11.9\,Gyr; see][]{Batalha:11},
whereas for background blends we have adopted for the binary a
representative 3\,Gyr isochrone of solar metallicity, although these
parameters have a minimal impact on the results.  For full details of
the technique we refer the reader to the references above. Three
recent changes and improvements that are especially relevant to the
application to \koi\ are described next:

(\emph{i}) The relatively long orbital period of \koi\ (45.3
days) precludes us from assuming that the eccentricity ($e$) is zero,
as we were able to suppose in previous applications of \blender\ to
Kepler-9\,d and Kepler-10\,b, which have periods of 1.59 and 0.84
days, respectively.  The reason this matters is that the duration of
the transit is set, among other factors, by the size of the secondary
star. Eccentricity can alter the speed of the tertiary around the
secondary, making it slower or faster than in the circular case
depending on the orientation of the orbit (longitude of periastron,
$\omega$).  Given a fixed (measured) duration, blends with smaller or
larger secondary stars than in the circular case may still provide
satisfactory fits to the light curve, effectively increasing the pool
of potential false positives.

\blender\ now takes this into account, although rather than using as
parameters $e$ and $\omega$, which are the natural variables employed
in the binary light-curve generating routine at the core of \blender\
\citep[see][]{Torres:11}, a more convenient variable that captures the
effects of both is the duration relative to a circular orbit.
Following \cite{Winn:10}, this may be expressed as $D/D_{\rm circ}
\approx \sqrt{1-e^2}/(1+e\sin\omega)$. Operationally, then, we vary
$D/D_{\rm circ}$ over wide ranges as we explore different blend
scenarios, and for each value we infer the corresponding values of $e$
and $\omega$.  In practice, in order to solve for \{$e$, $\omega$\}
from $D/D_{\rm circ}$ it is only necessary to consider the limiting
cases with $\omega = 90\arcdeg$ and 270\arcdeg, corresponding to
transits occurring at periastron and apastron, respectively, since
these are the orientations resulting in the minimum and maximum
durations for a given eccentricity. Other combinations of $e$ and
$\omega$ will lead to intermediate relative durations that are already
sampled in our $D/D_{\rm circ}$ grid. It is worth noting that use of
only these two values of $\omega$ leads to predicted secondary
eclipses in the simulated light curves that are always located at
phase 0.5, whereas secondary eclipses in the real data might be
present at any phase. For our purposes this is of no consequence, as
\koi\ has already had its light curve screened for secondary eclipses
at any phase that might betray a false positive, as part of the
vetting process. No such features are present down to the 100~ppm
level. Thus, any simulated light curves from \blender\ that display a
significant secondary eclipse will yield poor fits no matter where the
secondary eclipse happens to be, and will lead to the rejection of
that particular blend scenario.

(\emph{ii}) For each false positive configuration \blender\ can
predict the overall photometric color of the blend, for comparison
with the measured color index of the candidate as reported in the
\kepler\ Input Catalog \citep[KIC;][]{Brown:11}. A color index such as
$K\!p-K_s$, where $K\!p$ is the \kepler\ magnitude and $K_s$ derives
from the 2MASS catalog, provides a reasonable compromise between
wavelength leverage and the precision of the index. The latter varies
typically between 0.015 and 0.030 mag, depending on the passband and
the brightness of the star \citep[see][]{Brown:11}.  We consider a
particular blend to be rejected when its predicted color deviates from
the KIC value by more than three times the error of the latter. As it
turns out, color is a particularly effective way of rejecting blends
that include secondary stars of a different spectral type than the
primary, such as those that become possible when allowing for
eccentric orbits.

(\emph{iii}) Recent refinements in the resolution of the
\blender\ simulations to better explore parameter space, in addition
to the inclusion of eccentricity (or $D/D_{\rm circ}$) as an extra
variable, have increased the complexity of the problem as well as the
computing time (by nearly two orders of magnitude) compared to the
relatively simple case of circular orbits. The number of different
parameter combinations examined with \blender\ (and corresponding
light-curve fits) can approach $7 \times 10^8$ in some cases.
Consequently the simulations are now performed on the Pleiades cluster
at the NASA Advanced Supercomputing Division, located at the Ames
Research Center (California), typically on 1024 processors running in
parallel. For convenience hierarchical triple configurations (4
parameters) and background/foreground blends (5 parameters) are
studied separately, each for the two separate cases of stellar and
planetary tertiaries (for a total of four grids). One additional fit
is carried out using a true transiting planet model to provide a
reference for the quality of the false positive fits in the other
grids.

The discriminating value of the shape information contained in the
light curves, mentioned at the beginning of this section, is
highlighted by our \blender\ results for Kepler-10\,b, as described by
\cite{Batalha:11}. In that study it was found that \emph{all}
background eclipsing binary configurations with stellar tertiaries
yield very poor fits to the \kepler\ light curve, and are easily
rejected. The underlying reason is that all such blend models predict
obvious brightness changes out of eclipse (ellipsoidal variations)
with an amplitude that is not seen in the data, and that are a
consequence of the very short orbital period.
\footnote{Note that the present post-processing of \kepler\ data in
preparation for the \blender\ analyses (see
Sect.~\ref{sec:blender_app}) artificially suppresses out-of-eclipse
variations to some extent, typically by median filtering, so that the
light curves for periods as short as that of Kepler-10\,b (0.84 days)
are rendered essentially flat except for the transits themselves. In
this sense the situation is similar to that mentioned earlier
regarding the presence of secondary eclipses: obvious ellipsoidal
variability in the raw data would normally trigger a false positive
warning during the vetting process, preventing the target from
becoming an object of interest.  But if it reaches KOI status, we
assume that out-of-eclipse modulations are insignificant so that the
comparison with any \blender\ model in which those variations are
present is meaningful and would yield a poor fit, sufficient in most
cases to reject the blend.}
Hierarchical triple scenarios were also excluded based on joint
constraints from \blender\ and other follow-up observations. The only
configurations providing suitable alternatives to the true planet
scenario involved stars in the foreground or background of the target
that are orbited by a larger transiting planet. The considerable
reduction in the blend frequency from the exclusion of all background
eclipsing binaries led to a false alarm probability low enough to
validate Kepler-10\,b with a very high level of confidence,
\emph{independently} of any spectroscopic evidence.  This remarkable
result speaks to the power of \blender\ when combined with all other
observational constraints. It also assumes considerable significance
for Kepler-10\,b, given that it was not possible to provide separate
proof of the planetary nature of this signal in the \cite{Batalha:11}
study from an examination of the bisector spans. The scatter of the
bisector span measurements (10.5~\ms) was three times larger than the
RV semi-amplitude (3.3~\ms), rendering them inconclusive.

The situation regarding the \blender\ analysis of the \koi\ signal in
the \cite{Batalha:11} study was very different: the orbital period is
much longer, and ellipsoidal variations are predicted to be
negligible, so that background eclipsing binaries with stellar
tertiaries remain viable blends. This, and the added complication from
eccentric orbits, hindered the efforts of those authors to validate
this candidate. With the benefit of the enhancements in \blender\
described above, we are now in a better position to approach this
problem anew.

As follow-up observations provide important constraints that are
complementary to those supplied by \blender, and play an important
role in determining the false alarm rate for the planetary nature of
\koi\ (Sect.~\ref{sec:statistics}), we describe those first below,
beginning with our new near-infrared \spitzer\ observations.

\section{Observational constraints}
\label{sec:constraints}

\subsection{\wspitzer\ observations of \koi}
\label{sec:spitzer_obs}

\koi\ was observed during two transits with the IRAC instrument on the
\spitzer\ Space Telescope \citep{werner04,fazio04} at 4.5\,\micron\
(program ID 60028).  The observations were obtained on UT 2010 August
30 and November 15, with each visit lasting approximately 15\,hr
10\,min.  The data were gathered in full-frame mode ($256\times256$
pixels) with an exposure time of 6.0\,s per image, which resulted in
approximately a 7.1\,s cadence and yielded 7700 images per visit.

The method we used to produce photometric time series from the images
is described by \cite{desert09}.  It consists of finding the centroid
position of the stellar point spread function (PSF) and performing
aperture photometry using a circular aperture on individual exposures.
The images used are the Basic Calibrated Data (BCD) delivered by the
\emph{Spitzer} archive.  These files are corrected for dark current,
flat-fielding, and detector non-linearity, and are converted to flux
units.  We converted the pixel intensities to electrons using the
information on the detector gain and exposure time provided in the
FITS headers.  This facilitates the evaluation of the photometric
errors.  We extracted the UTC-based Julian date for each image from
the FITS header (keyword DATE\_OBS) and corrected to mid-exposure.  We
converted to TDB-based barycentric Julian dates using the
\texttt{UTC2BJD}\footnote{{\tt
http://astroutils.astronomy.ohio-state.edu/time/}} procedure developed
by \citet{eastman10}.  This program uses the JPL Horizons ephemeris to
estimate the position of the \spitzer\ spacecraft during the
observations.  We then corrected for transient pixels in each
individual image using a 20-point sliding median filter of the pixel
intensity versus time.  To do so, we compared each pixel's intensity
to the median of the 10 preceding and 10 following exposures at the
same pixel position, and we replaced outliers greater than $4\sigma$
with their median value.  The fraction of all pixels we corrected is
0.02\% for the first visit and 0.06\% for the second.

\begin{figure}[b!]
\vskip 10pt
\begin{center}
\epsscale{1.2} 
\plotone{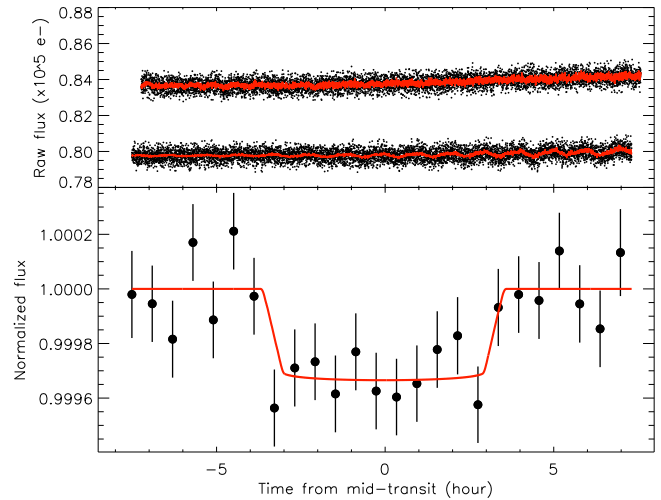}
%
\vskip -10pt
 \caption{\spitzer\ transit light-curves of \koi\ observed in the IRAC
 band-pass at 4.5\,\micron. {\it Top:} Raw measurements (black points)
 with the same data binned by two superimposed (12\,s bins, red
 points).  {\it Bottom:} Measurements combined from the two visits and
 binned in 36\,min bins (295 points per bin), along with the best-fit
 limb-darkened transit model (integrated over the same duration). Both
 the data and the model shown here have been corrected for
 instrumental errors.\label{fig:spitzerlightcurves}}
\end{center}
\end{figure}

The centroid position of the stellar PSF was determined using the
DAOPHOT-related procedures \texttt{GCNTRD}, from the IDL Astronomy
Library\footnote{{\tt
http://idlastro.gsfc.nasa.gov/homepage.html}}. We applied the
\texttt{APER} routine to perform aperture photometry with a circular
aperture of variable radius, using a range of radii between 1.5 and 8
pixels in steps of 0.5.  The propagated uncertainties were derived as
a function of the aperture radius, and we adopted the aperture
providing the smallest errors.  We found that the transit depths and
errors varied only weakly with aperture radius for all light-curves
analyzed in this project.  The optimal aperture was found to have a
radius of 4.0 pixels.  

We estimated the background by examining a histogram of counts from
the full array. We fit a Gaussian curve to the central region of this
distribution (ignoring bins with high counts, which correspond to
pixels containing stars), and we adopted the center of this Gaussian
as the value of the residual background intensity.  As seen already in
previous \wspitzer\ observations \citep{deming11, beerer11}, we found
that the background varies by 20\% between three distinct levels from
image to image, and displays a ramp-like behavior as function of time.
The contribution of the background to the total flux from the stars is
low for both observations, from 0.1\% to 0.55\% depending on the
image.  Therefore, photometric errors are not dominated by
fluctuations in the background.  We used a sliding median filter to
select and trim outliers in flux and position greater than 5$\sigma$,
representing 1.6\% and 1.3\% of the data for the first and second
visits, respectively.  We also discarded the first half-hour's worth
of observations, which is affected by significant telescope jitter
before stabilization.  The final number of photometric measurements
used is 7277 and 7362.

The raw time series are presented in the top panel of
Figure~\ref{fig:spitzerlightcurves}. We find that the point-to-point
scatter in the photometry gives a typical signal-to-noise ratio (S/N)
of 280 per image, which corresponds to 90\% of the theoretical
signal-to-noise.  Therefore, the noise is dominated by Poisson
statistics.

\subsection{Analysis of the \wspitzer\ light curves, and results}
\label{sec:spitzer_analysis}

In order to determine the transit parameters and associated
uncertainties from the \spitzer\ time series we used a transit light
curve model multiplied by instrumental decorrelation functions, as
described by \cite{desert11a}.  The transit light curves were computed
with the IDL transit routine \texttt{OCCULTSMALL} from
\cite{mandel02}.  For the present case we allowed for a single free
parameter in the model, which is the planet-to-star radius ratio
$R_p/R_\star$ (or equivalently, the depth, in the absence of limb
darkening).  The normalized orbital semi-major axis (system scale)
$a/R_\star$, the impact parameter $b$, the period $P$, and the time of
mid transit $T_c$ were held fixed at the values derived from the
\kepler\ light curve, as reported by \cite{Batalha:11} and summarized
below in Sect.~\ref{sec:discussion}.  Limb darkening is small at
4.5\,\micron, but was nevertheless included in our modeling using the
4-parameter law by \cite{Claret:00} and theoretical coefficients
published by \cite{Sing:10}.

The \spitzer/IRAC photometry is known to be systematically affected by
the so-called ``pixel-phase effect'' \citep[see,
e.g.,][]{charbonneau05,knutson08}.  This effect is seen as
oscillations in the measured fluxes with a period corresponding to
that of the telescope pointing jitter. For the first visit this period
was 70~min, and the amplitude of the oscillations was approximately
2\% peak-to-peak; for the second visit the period was 35~min, and the
amplitude about 1\%.  We decorrelated our signal in each channel using
a linear function of time for the baseline (two parameters) and a
quadratic function of the PSF position (four parameters) to correct
the data for each channel.  We performed a simultaneous
Levenberg-Marquardt least-squares fit to the data \citep{markwardt09}
to determine the transit and instrumental model parameters (7 in
total).  The errors on each photometric point were assumed to be
identical, and were set to the rms residual of the initial best fit.
To obtain an estimate of the correlated and systematic errors in our
measurements \citep{pont06} we used the residual permutation bootstrap
technique, or ``Prayer Bead'' method, as described by
\citet{desert09}.  In this method the residuals of the initial fit are
shifted systematically and sequentially by one frame, and then added
to the transit light curve model before fitting again.  We considered
asymmetric error bars spanning 34\% of the points above and below the
median of the distributions to derive the $1\sigma$ uncertainties for
each parameter, as described by \citet{desert11b}.

The bottom panel of Figure~\ref{fig:spitzerlightcurves} shows the
best-fit model superimposed on the observations from the two visits
combined, with the data binned in 36\,min bins for clarity (295 points
per bin). The transit depths at 4.5\,\micron\ (after removing
limb-darkening effects) are $353^{+115}_{-133}$ ppm for the first
visit and $339^{+85}_{-110}$ for the second, which are in good
agreement with each other.  The weighted average depth of $344 \pm 85$
is consistent with the non-limb-darkened value of $376 \pm 9$ ppm
derived from the \kepler\ light curve \citep{Batalha:11} well within
the 1$\sigma$ errors, strongly suggesting the transit is achromatic,
as expected for a planet.

The above \spitzer\ observations provide a useful constraint on the
kinds of false positives (blends) that may be mimicking the \koi\
signal. For example, if Kepler-10 were blended with a faint unresolved
background eclipsing binary of much later spectral type that manages
to reproduce the transit depth in the \kepler\ passband, the predicted
depth at 4.5\,\micron\ may be expected to be larger because of the
higher flux of the contaminating binary at longer wavelengths compared
to Kepler-10. Since the transit depth we measure in the near infrared
is about the same as in the optical, this argues against blends
composed of stars of much later spectral type. Based on model
isochrones and the properties of the target star (see below), we
determine an upper limit to the secondary masses of
0.77\,$M_{\sun}$. This \spitzer\ constraint is used in
Sect.~\ref{sec:blender_app} to eliminate many blends.

\subsection{Additional observational constraints on possible false positives}
\label{sec:followup}

Further constraints of a different kind are provided by
high-resolution imaging as described in more detail by
\cite{Batalha:11}. Briefly, these consist of speckle observations
obtained on UT 2010 June 18 with a two-color (approximately $V$ and
$R$) speckle camera on the WIYN 3.5\,m telescope on Kitt Peak
\citep[see][]{Howell:11}, and near-infrared ($J$-band) adaptive optics
(AO) observations conducted on UT 2009 September 8 with the PHARO
camera on the 5\,m Palomar telescope. No companions were detected
around Kepler-10 within 1\farcs5 (for speckle) or 12\farcs5 (AO), and
more generally these observations place strong limits on the presence
of other stars as a function of angular separation (down to 0\farcs05
in the case of speckle) and relative brightness (companions as faint
as $\Delta J = 9.5$ for AO). These sensitivity curves are shown in
Fig.~9 of \cite{Batalha:11}, and we make use of that information
below.

High-resolution spectra described also by \cite{Batalha:11} and
obtained with the HIRES instrument on the 10\,m Keck~I telescope place
additional limits on the presence of close companions falling within
the spectrograph slit (0\farcs87), such that stars within about 2
magnitudes of the target would generally have been seen. A small
chance remains that these companions could escape detection if their
radial velocity happens to be within a few \kms\ of that of the target
\citep[which is a narrow-lined, slowly rotating star with $v \sin i =
0.5 \pm 0.5\,\kms$;][]{Batalha:11}, so that the spectral lines are
completely blended. This would be extremely unlikely for a chance
alignment with a background/foreground star, but not necessarily for
physically associated companions in wide orbits, i.e., with slow
orbital motions. We explored this through Monte Carlo simulations. The
results indicate that the probability of having a physical companion
within a conservative range of $\pm$10\,\kms\ of the RV of the target
that would also go unnoticed in our speckle observations, and that
additionally would not induce a RV drift on the target large enough to
have been detected in the high-precision measurements of
\cite{Batalha:11}, is only about 0.1\%.

Finally, an analysis of the image centroids measured from the \kepler\
observations rules out background objects of any brightness beyond
about 2\arcsec\ of the target. This exclusion limit (equivalent to
half a pixel) is considerably more conservative than the 0\farcs6
reported by \cite{Batalha:11}, and accounts for saturation effects not
considered earlier (given that at $K\!p = 10.96$ the star is very
bright by \kepler\ standards) as well as quarter-to-quarter variations
(where ``quarters'' usually represent 3-month observing blocks
interrupted by spacecraft rolls required to maintain the proper
illumination of the solar panels).

\section{Application of \blender\ to \koi}
\label{sec:blender_app}

The \kepler\ photometry used here is the same as employed in the work
of \cite{Batalha:11}, and was collected between 2009 May 2 and 2010
January 9. These dates correspond to \kepler\ Quarter~0 (first nine
days of commissioning data) through the first month of Quarter~4.  For
this study we used only the long-cadence observations (10,870
measurements) obtained by the spacecraft at regular intervals of about
29.4 min. All blend models generated with \blender\ were integrated
over this time interval for comparison with the measurements. The
original data have been de-trended for this work by removing a
first-order polynomial, and then applying median filtering with a
2-day wide sliding window. Observations that occur during transits
were masked and did not contribute to the median calculation. Because
this sliding window is considerably shorter than the 45.3-day orbital
period, any ellipsoidal variations present in the original data should
be largely preserved, although in any case they are expected to be
very small for binaries with periods as long as this.  We adopted also
the ephemeris of mid-transit for \koi\ as reported by
\cite{Batalha:11}, which is $T_c~{\rm [BJD]} = 2,\!454,\!971.6761 + N
\times 45.29485$ days, where $N$ is the number of cycles from the
reference epoch.

Because it is relatively bright ($K\!p = 10.96$), Kepler-10 was also
observed by the Mission with a shorter cadence of approximately 1~min
for a period of several months to allow an asteroseismic
characterization of the star.  A total of 19 oscillation frequencies
were detected, and enabled a very precise determination of the mean
stellar density. When combined with stellar evolution models and a
spectroscopic determination of the effective temperature and chemical
composition, the resulting parameters for the star are very well
determined. Kepler-10 is relatively old ($> 7.4$\,Gyr) but is
otherwise quite similar to the Sun, with a temperature of $T_{\rm eff}
= 5627 \pm 44$\,K, a mass and radius of $M_{\star} = 0.895 \pm
0.060$\,$M_{\sun}$ and $R_{\star} = 1.056 \pm 0.021$\,$R_{\sun}$, and
a composition [Fe/H] $= -0.15 \pm 0.06$ slightly below solar
\citep{Batalha:11}.

As indicated earlier we considered four general scenarios for false
positives: chance alignments (a pair of background/foreground
eclipsing objects) and hierarchical triple systems, each with
tertiaries that can be either stars or planets.  The free parameters
were varied over the following ranges: secondary mass $M_2$ between
0.10 and 1.40\,$M_{\sun}$, in steps of 0.02\,$M_{\sun}$; tertiary mass
$M_3$ between 0.10 and $M_2$, also in steps of 0.02\,$M_{\sun}$;
tertiary radius $R_3$ between 0.06 and 2.00\,$R_{\rm Jup}$ in steps of
0.02\,$R_{\rm Jup}$; impact parameter $b$ between 0.00 and 1.00 in
steps of 0.05; relative duration $D/D_{\rm circ}$ between 0.2 and 4.6
in steps of 0.2, corresponding to eccentricities up to 0.92 and values
of $\omega$ of 90\arcdeg\ and 270\arcdeg\ (see
Sect.~\ref{sec:blender}); and relative distance $\Delta\delta$ (distance
modulus difference) between $-5.0$ and $+9.0$ in steps of 0.5 mag,
except for hierarchical triple configurations, for which $\Delta\delta
= 0$.

The goodness of the fit of each of the large number of synthetic light
curves generated by \blender\ is quantified here by computing the
$\chi^2$ statistic and comparing it with that of the best planet model
fit. The difference can be assigned a significance level (or false
alarm rate) that depends on the number of free parameters of the
problem. For example, for a blend scenario corresponding to a
hierarchical triple system (4 degrees of freedom), a trial model
giving a worse fit than the planet solution by $\Delta\chi^2 =$ 4.72
is statistically different at the 1$\sigma$ level, assuming Gaussian
errors \citep[see, e.g.,][]{Press:92}.  A fit that is worse by
$\Delta\chi^2 =$ 16.3 is different at the 3$\sigma$
level. Hierarchical triple blends giving poorer fits than this are
considered here to be ruled out by the \kepler\ photometry. For
background/foreground scenarios (5 degrees of freedom) the 3$\sigma$
blend rejection level is $\Delta\chi^2 =$ 18.2.

\subsection{\blender\ results}
\label{sec:blender_results}

In this section we describe the simulations carried out for the four
general blend configurations mentioned above.  Although the
secondaries for the background scenarios can in principle also be
evolved stars (giants), as opposed to main-sequence stars, we
consistently found that the transit light curves generated by such
systems give a very poor match to the observations because they do not
have the right shape (the ingress/egress phases are too
long). Therefore, we restricted our exploration of parameter space to
main-sequence stars only.

An additional possibility for a false positive may stem from an error
in the determination of the orbital period. If the true period were
twice the nominal value, alternating transit events would correspond
to primary and secondary eclipses, implicating a blended eclipsing
binary. The primary and secondary eclipses would often (but not
always) be of different depth. As part of the vetting process for each
candidate, the \kepler\ Team examines the even-numbered and
odd-numbered events to look for differences in depth that may indicate
a false positive of this kind. As described by \cite{Batalha:11}, no
significant differences were found for \koi\ beyond the 2$\sigma$
level, where $\sigma$ represents the uncertainty in the transit depth
(9~ppm). Nevertheless, as the possibility still exists that the
components of the eclipsing binary are identical, experiments were run
with \blender\ to examine the transit shape produced by such
scenarios, and it was found that the ingress and egress phases are
always much too long compared to the observations, as expected for two
equal-size stars eclipsing each other. Thus, these scenarios are
easily ruled out as well.

\subsubsection{Background eclipsing binaries (star\,+\,star)}
\label{sec:bs}

\begin{figure}[h!]
\epsscale{1.15}
\plotone{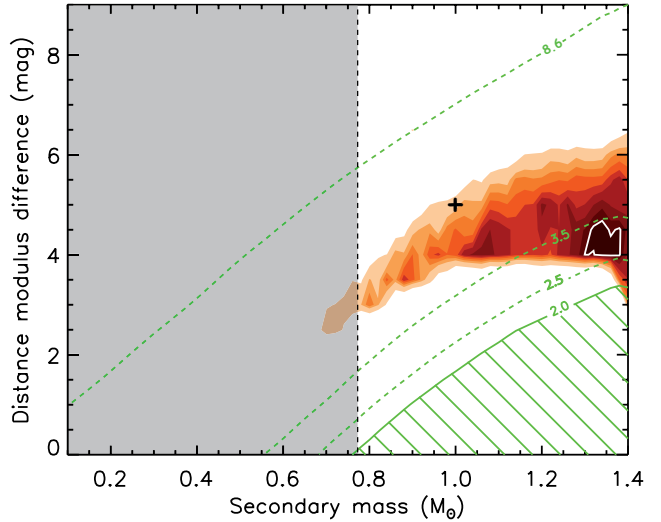}
\figcaption[]{Map of the $\chi^2$ surface (goodness of fit)
corresponding to a grid of blend models for KOI-072.02 involving
background eclipsing binaries. The linear separation between the
binary and the primary is cast in terms of the distance modulus
difference. Contours are drawn as a function of the $\chi^2$
difference from the best planet model fit (expressed in units of the
significance level of the difference, $\sigma$), and are plotted here
as a function of the mass of the secondary star.  Only blends within
the small white contour yield acceptable fits to the light curve
(within $3 \sigma$ of the planet fit). Other colored areas correspond
to regions of parameter space giving increasingly worse fits
(4$\sigma$, 5$\sigma$, etc.), representing blends we consider to be
ruled out. The \spitzer\ constraint is indicated by the shaded area:
blends with secondary masses in this region are excluded (see
Sect.~\ref{sec:spitzer_analysis}), although \blender\ itself already
rules out all of these scenarios.  Green lines running diagonally from
the lower left to the upper right are labeled with the magnitude
difference $\Delta K\!p$ of the blended binary relative to the target
star. The hatched region below the $\Delta K\!p = 2$ mag line
represents blends with secondary stars bright enough that they would
generally be detected in our spectroscopy. Viable blends within the
3$\sigma$ contour are seen to be confined to a narrow range of
magnitude differences ($2.5 \leq \Delta K\!p \leq 3.5$, dashed green
lines). The dashed line at $\Delta K\!p = 8.6$ indicates the envelope
for the faintest blends that would be capable of reproducing the
measured depth based on simple-minded estimates described in the
text. As seen, \blender\ provides much tighter constraints than this.
The cross corresponds to a blend model that gives the fit shown in the
bottom panel of Figure~\ref{fig:fits}.
\label{fig:backstar}}
\end{figure}

\begin{figure}[h!]
\epsscale{1.15}
\plotone{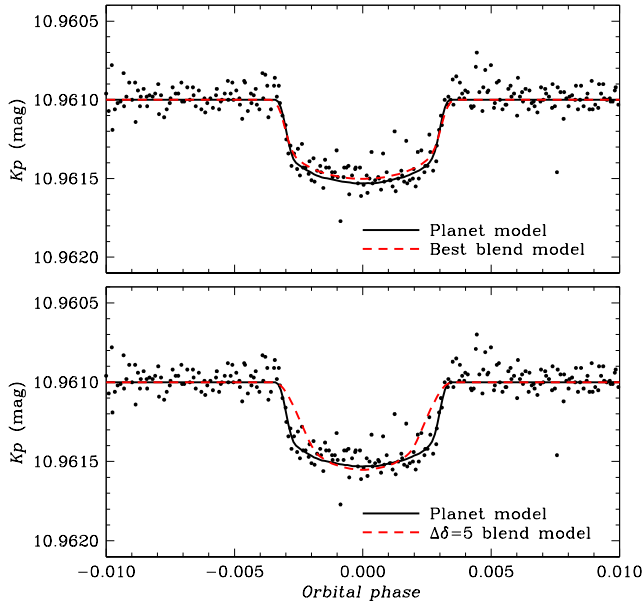}
\figcaption[]{Long-cadence \kepler\ observations for \koi\ compared
with two different blend models involving background eclipsing
binaries (red lines), and shown against the best fit planet model for
reference (black line). All models are integrated over the duration of
one cadence (29.4 min). {\it Top:} The blend model shown is the one
giving the best fit for this type of scenario (2.1$\sigma$ difference
compared to the planet fit). {\it Bottom:} Example of a blend model
(indicated with a cross in Figure~\ref{fig:backstar}) that illustrates
the use of shape information by \blender\ in a case that would naively
be expected to be a viable false positive scenario (see text). This
particular scenario corresponds to a secondary mass of $M_2 =
1.0\,M_{\sun}$ and a distance modulus difference of 5~mag relative to
the target, giving a brightness difference in the $K\!p$ band of 5.6
mag. Although it matches the depth and total duration of the transit,
the ingress and egress phases are not well reproduced, so that the
overall quality of the fit is poor and the blend is ruled out at more
than the 10-$\sigma$ level.\label{fig:fits}}
\end{figure}

The simulations with \blender\ indicate that few background blend
scenarios with stellar tertiaries are able to mimic the transit
features in the light curve at an acceptable level, and they all
correspond to somewhat eccentric orbits.  In Figure~\ref{fig:backstar}
we show the goodness of fit of these scenarios, with the small closed
3$\sigma$ contour representing the region of parameter space within
which the fits are satisfactory, according to the criteria given
above. Only blends with secondary masses $M_2$ larger than about
1.3\,$M_{\sun}$ are allowed, and the eclipsing binary can only be
within a small range of distances behind the target ($4.0 \lesssim
\Delta\delta \lesssim 4.7$) for the dilution effect to be just right,
such that the corresponding apparent brightness difference $\Delta
K\!p$ is between 2.5 and 3.5 mag (see figure). The best among these
blend models (located near the center of the contour) provides a fit
that is about 2.1$\sigma$ worse than a planet model (but still
acceptable), and is shown in the top panel of Figure~\ref{fig:fits}
compared against the planet model. The tertiary stars in these blends
are constrained to be very small, between 0.10 and 0.16\,$M_{\sun}$.

That most blends involving background eclipsing binaries can be ruled
out may appear somewhat surprising, and is worth investigating.
Indeed, for a given measured transit depth $d_{\rm tran}$, a blend can
only reproduce the light curve if it contributes at least a fraction
$d_{\rm tran}$ of the total flux collected in the \kepler\ aperture.
Thus, one would expect that binaries as faint as $\Delta K\!p = -2.5
\log(d_{\rm tran}) \approx 8.6$ mag relative to the target should be
able to match that amount of dimming if they were totally eclipsed
\citep[see, e.g.,][]{Morton:11}, and furthermore, that the measured
duration could also be reproduced by a large range of secondary sizes
with an appropriate combination of orbital eccentricity and
$\omega$. Yet we find that no blends fainter than $\Delta K\!p = 3.5$
give tolerable fits to the light curve (see
Figure~\ref{fig:backstar}).  A visual understanding of the underlying
reason for this may be seen in the bottom panel of
Figure~\ref{fig:fits}, in which we show a blend model that one would
naively expect should be able to match the observations, according to
the crude recipe described above. This particular blend scenario is
marked with a cross in Figure~\ref{fig:backstar}, and corresponds to
$\Delta\delta = 5$ and $M_2 = 1.0$\,$M_{\sun}$, resulting in a
magnitude difference of $\Delta K\!p = 5.6$ for the EB relative to the
target. While this model does yield a good match to the measured
depth, and even the total duration, it doesn't perform nearly as well
in the ingress/egress phases, which are too long when compared against
the observations. The quality of this fit relative to the best planet
fit, which can also be seen in the figure, corresponds to a
10.1$\sigma$ difference, and therefore \blender\ rejects it. Thus, the
reason most blends of this class can be ruled out is ultimately the
high precision of the \kepler\ light curves, which provides a very
strong constraint on the shape of the transit light curve, and in
particular on the size ratio between the secondary and tertiary, which
sets the duration of the ingress and egress phases.

\subsubsection{Background/foreground star\,+\,planet pairs}
\label{sec:beb1}

\begin{figure}
\epsscale{1.15}
\plotone{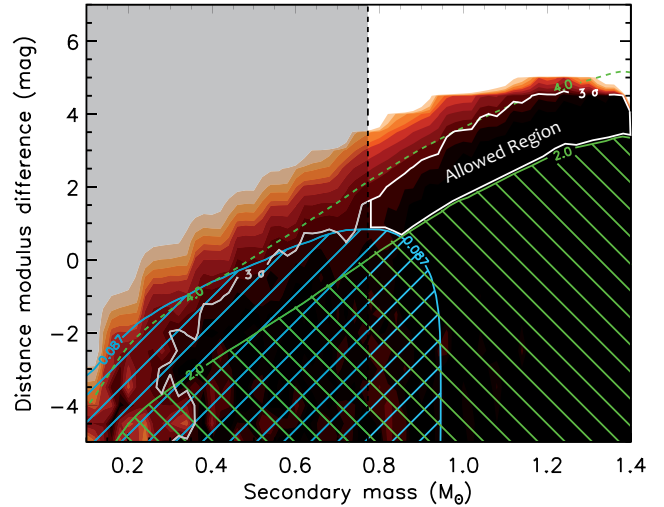}
\figcaption[]{Similar to Figure~\ref{fig:backstar}, for blends
involving background systems consisting of a star transited by a
planet. The color scheme is the same as in Figure~\ref{fig:backstar}.
Blends giving fits no worse than 3$\sigma$ from the best planet fit
are below the thick white contour labeled with that confidence level.
The shaded left-hand side of the diagram corresponds to secondary
masses excluded by constraints from the \spitzer\ observations.  Lines
of constant magnitude difference relative to the target are shown in
green, running diagonally from the lower left to the upper right.  The
dashed one at the top represents the boundary for the faintest viable
blends (tangent to the white 3$\sigma$ contour). The solid green line
below and parallel to it ($\Delta K\!p = 2$ mag) and hatched region to
the right marks the area of parameter space excluded by our
spectroscopic constraints. The blue curve and hatched region to the
left represent blends that are excluded because they are too red in
$K\!p - K_s$ compared to the target. Note that the colors of the
blended stars are computed from a different isochrone than that of the
target, which explains why blends with secondaries of the same mass as
the target are ruled out for being too red. The combination of all
these constraints leaves only a reduced area of parameter space
(labeled ``Allowed Region'') where blend models give tolerable fits to
the \kepler\ light curve and are not ruled out by any of our follow-up
observations.
\label{fig:back_plan}}

\end{figure}

There is a very broad range of blends consisting of a background or
foreground star transited by a planet (as opposed to a star) that are
found by \blender\ to give satisfactory fits to the data, as shown in
Figure~\ref{fig:back_plan}.  These viable blends occupy the area below
the 3$\sigma$ contour represented with a thick white line.  Secondary
stars of all spectral types (masses) are permitted, in principle,
although in practice other constraints described below eliminate a
substantial fraction of them.  All of these blends involve
secondary+tertiary pairs that are within 4 magnitudes of the target in
the \kepler\ passband (diagonal dashed line in the figure). The
tertiary sizes in these blends range from 0.42\,$R_{\rm Jup}$ to
1.84\,$R_{\rm Jup}$.

Our \wspitzer\ observations set a lower limit of about
0.77\,$M_{\sun}$ for the secondary masses of these blends, as
described earlier; scenarios involving redder stars would result in
transits at 4.5\,\micron\ significantly deeper than we observe (i.e.,
deeper than the measured depth + 3$\sigma$).  This exclusion region is
indicated by the shaded area.  Additionally, blends that are much
brighter than $\Delta K\!p = 2$ would most likely have been detected
spectroscopically \citep[see][]{Batalha:11}, so we consider those to
be ruled out as well.  We indicate this with the green hatched region
in the lower right-hand side of the figure.  Finally, the colors of
the background/foreground configurations simulated with \blender\
provide a further constraint which is represented by the blue hatched
area on the lower left of the figure. This swath of parameter space is
excluded because the blends are significantly redder than the color
index measured for Kepler-10 ($K\!p-K_s = 1.465 \pm 0.029$), by more
than three times the uncertainty in the observed index. As a result of
these complementary constraints, the only section of parameter space
remaining for viable blends involving star+planet pairs is the area
under the 3$\sigma$ contour and limited from below and on the left by
the hatched areas (color and brightness conditions) and shaded area
(\spitzer\ constraint), respectively. All of these blends have the
eclipsing pair behind the target (foreground scenarios are all ruled
out).

We note that in this star+planet blend scenario white dwarfs can also
act as tertiaries, as long as they are cooler than the secondaries so
that they do not lead to deep occultation events that would have been
seen in the light curve of \koi.  The above range of tertiary radii
(0.42\,$R_{\rm Jup}$ to 1.84\,$R_{\rm Jup}$) excludes essentially all
cool carbon-oxygen and oxygen-neon white dwarfs more massive than
about 0.4\,$M_{\sun}$, as these are smaller than the lower limit set
by \blender, which corresponds to 4.7\,$R_{\earth}$ \citep[see,
e.g.,][]{Panei:00}.  Low-mass helium-core or oxygen-core white dwarfs
that are the product of common-envelope evolution in binary stars can
be considerably larger in size, although they appear to be very
rare. The \kepler\ Mission itself has uncovered only three examples to
date \citep{Rowe:10, Carter:11}.  However, all of them are very hot
($T_{\rm eff} > 10,000$\,K), and produce deep and unmistakable
flat-bottomed occultation signals. Model calculations such as those of
\cite{Panei:07} show that as these helium-core white dwarfs cool,
their radii quickly become Earth-size or smaller.  Therefore, we do
not consider white dwarfs to be a significant source of blends for
\koi.

\subsubsection{Hierarchical triple scenarios (star\,+\,star and star\,+\,planet blends)}
\label{sec:beb2}

\begin{figure}
\epsscale{1.15}
\plotone{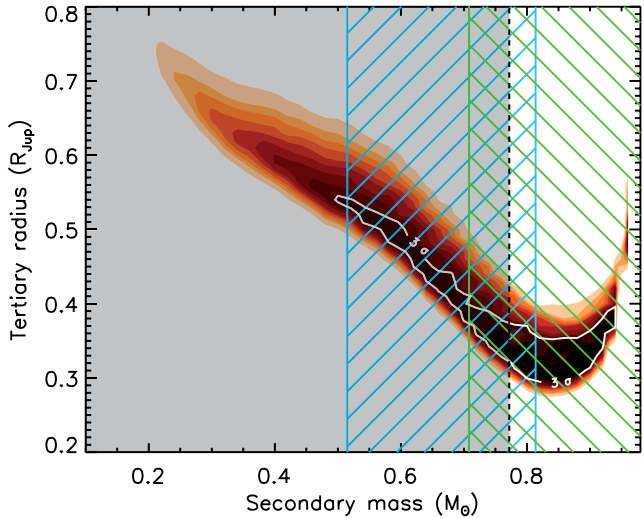}
\figcaption[]{Similar to Figure~\ref{fig:backstar}, for the case of
hierarchical triple systems in which the secondary star is transited
by a planet. The color scheme is the same as in
Figure~\ref{fig:backstar}. In this case the vertical axis shows the
tertiary sizes.  The constraints from \spitzer\ (gray shaded area to
the left of 0.77\,$M_{\sun}$), color information (blue hatched area on
the left), and spectroscopy (green hatched area on the right) are
shown as in previous figures. When taken together these constraints
eliminate all blends of this kind.\label{fig:htp_r3}}
\end{figure}

\begin{figure}
\epsscale{1.15}
\plotone{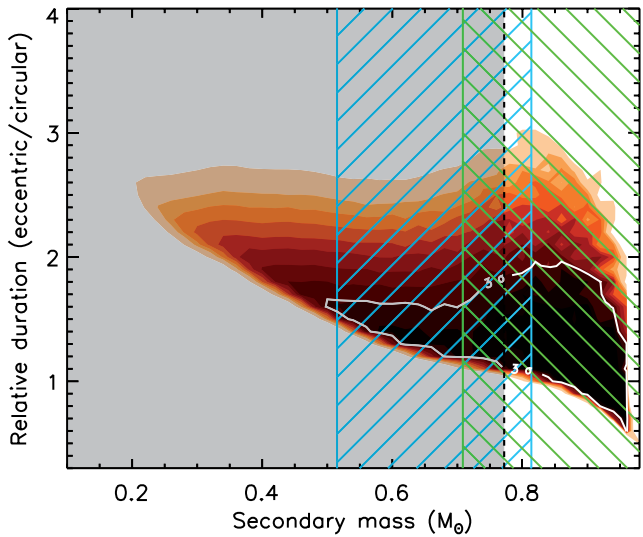}
\figcaption[]{Same as Figure~\ref{fig:htp_r3}, but with the vertical
axis showing the relative transit durations ($D/D_{\rm
circ}$).\label{fig:htp_dur}}
\end{figure}

Eclipsing binaries composed of two stars physically associated with
the target are clearly ruled out by \blender, as they produce very
poor fits to the \kepler\ light curves. For cases in which the
tertiaries are planets, viable scenarios identified by \blender\ span
a range of secondary masses and tertiary radii within the 3$\sigma$
contour shown in Figure~\ref{fig:htp_r3}. Most of these configurations
turn out to involve eccentric orbits, with transit durations longer
than those corresponding to circular orbits along with secondary stars
that are smaller than the primary (see Figure~\ref{fig:htp_dur}). Once
again other observational constraints are very complementary, and in
this case they are sufficient to exclude all of these blends.  For
example, the shaded area of parameter space to the left of
0.77\,$M_{\sun}$ is eliminated by the \spitzer\ observations, as
described earlier. The constraint on the $K\!p - K_s$ color (hatched
area on the left) is partly redundant with the NIR observations, but
extends to slightly larger secondary masses. And finally, the
spectroscopic constraint removes the remaining scenarios corresponding
to higher-mass (brighter) secondaries.

We conclude that of all the hierarchical triple blend scenarios that
are capable of precisely reproducing the detailed shape of the
\kepler\ transit light curve, \emph{none} would have escaped detection
by one or more of our follow-up efforts, including NIR \spitzer\
observations, high-resolution spectroscopy, or absolute photometry
(colors).\footnote{The possibility, remote as it may be, that the
target has a physically associated companion that is nearly of the
same brightness and that has managed to elude detection is always
present (see Sect.~\ref{sec:followup}), not only here but in all
previously discovered transiting planets.  For the present purposes we
do not consider this ``twin star'' scenario as a false positive in the
strict sense \citep[see also][]{Torres:11}, as the transiting object
would still be a planet, only that it would be larger than we thought
by about a factor of $\sqrt{2}$ because of the extra dilution from the
companion.\label{twin}} This highlights the importance of these types
of constraints for validating \kepler\ candidates, given that blends
involving physically associated stars would generally be spatially
unresolved by our high-resolution imaging with adaptive optics or
speckle interferometry, and they would typically also be below the
sensitivity limits of our centroid motion analysis, so that they would
not be detected by those means. Therefore, the only blends we need to
be concerned about for \koi\ are those consisting of stars in the
background of the target that are orbited by other stars or by
transiting planets.

\section{{\it A priori} likelihood of remaining blend scenarios for KOI-72.02}
\label{sec:blendfreq}

In order to estimate the frequency of the blend scenarios (i.e.,
background configurations) that remain possible after applying
\blender\ and all other observational constraints, we follow a
procedure similar to that described by \cite{Torres:11} for
Kepler-9\,d. We appeal to the Besan\c{c}on Galactic structure models
of \cite{Robin:03} to predict the number density of background stars
of each spectral type (mass) and brightness around Kepler-10, in
half-magnitude bins, and we make use of estimates of the frequencies
of transiting planets and of eclipsing binaries from recent studies by
the \kepler\ Team to infer the number density of blends. Using
constraints from our high-resolution imaging \citep[specifically, the
sensitivity curves presented by][their Fig.~9]{Batalha:11} we
calculate the area around the target within which blends would go
undetected, and with this the expected number of blends.\footnote{In
the case of \koi\ the exclusion radius from the centroid motion
analysis, 2\arcsec\ (Sect.~\ref{sec:followup}), is significantly less
constraining than the high-resolution imaging, so is not as useful
here as it was for Kepler-9\,d.}

The recent release by \cite{Borucki:11b} of a list of 1235 candidate
transiting planets (KOIs) from \kepler\ provides a means to estimate
planet frequencies needed for our calculations, with significant
advantages over the calculations of \cite{Torres:11} for Kepler-9\,d,
which were based on the earlier list of candidates published by
\cite{Borucki:11a}. Not only is the sample now much larger, but the
knowledge of the rate of false positives for \kepler\ is also much
improved, and that rate is believed to be relatively small (20--40\%
depending on the level of vetting of the candidate, according to
\citealt{Borucki:11a}; less than 10\% according to the recent study by
\citealt{Morton:11}).  Thus, our results will not be significantly
affected by the assumption that all of the candidates are planets (see
also below). An additional assumption we make is that this census is
largely complete. Among these candidates we count a total of 267
having radii in the range allowed by \blender\ for the tertiaries of
viable blends (i.e., between 0.42 and 1.84\,$R_{\rm Jup}$). With the
total number of \kepler\ targets being 156,453 \citep{Borucki:11b},
the relevant frequency of transiting planets for our blend calculation
is $f_{\rm planet} = 267/156,\!453 = 0.0017$. \cite{Slawson:11} have
recently published a catalog of the 2165 eclipsing binaries found in
the \kepler\ field, from the first four months of observation. Only
the 1225 detached systems among these are considered here, since
binaries in the category of semi-detached, over-contact, or
ellipsoidal variables would not produce light curves with a shape
consistent with a transit. The frequency of eclipsing binaries for our
purposes is then $f_{\rm EB} = 1225/156,\!453 = 0.0078$.

Table~\ref{tab:new_stats} presents the results of our calculation of
the frequency of blends, separately for background blends with stellar
tertiaries (eclipsing binaries) and with planetary
tertiaries. Columns~1 and 2 give the $K\!p$ magnitude range of each bin
and the magnitude difference $\Delta K\!p$ relative to the target,
calculated at the upper edge of each bin. Column~3 reports the mean
number density of stars per square degree obtained from the
Besan\c{c}on models, for stars in the mass range allowed by \blender\
as shown in Figure~\ref{fig:backstar}. In column~4 we list the maximum
angular separation $\rho_{\rm max}$ at which stars in the
corresponding magnitude bin would go undetected in our imaging
observations, taken from the information in the work of
\cite{Batalha:11}. The product of the area implied by this radius and
the stellar densities in the previous column give the number of stars
in the appropriate mass range, listed in column~6 in units of
$10^{-6}$. Multiplying these figures by the frequency of eclipsing
binaries $f_{\rm EB}$ then gives the number of background star+star
blends in column~7. A similar calculation for the background
star+planet blends, making use of $f_{\rm planet}$, is presented in
columns~7--10. We sum up the contributions from each magnitude bin at
the bottom of columns~6 and 10.  The total number of blends we expect
{\it a priori} (blend frequency) is given in the last line of the
table by adding these two values together, and is ${\rm BF} = 1.62
\times 10^{-8}$. The calculations show that background blends
consisting of star+planet pairs contribute to this frequency about
three times more than background eclipsing binaries.

While we have assumed up to now that any companions to \koi\ within
$\Delta K\!p = 2$ mag of the target would have been seen
spectroscopically, we note that relaxing this condition to a much more
conservative $\Delta K\!p = 1$ has no effect at all on the contribution
from eclipsing binaries, and a negligible effect on the contribution
of star+planet scenarios.

\section{Likelihood of the planet interpretation for KOI-72.02}
\label{sec:statistics}

To obtain a Bayesian estimate of the probability that \koi\ is indeed
a planet as opposed to a false positive (or equivalently, the ``false
alarm rate'', FAR) we follow the general methodology of
\cite{Torres:11} and compare the {\it a priori} likelihoods of blends
and of planets: FAR = BF/PF. If the {\it a priori} blend frequency is
sufficiently small compared the planet frequency (PF), we consider the
planet validated.  Our {\it a priori} blend frequencies above
correspond to false positive scenarios giving fits to the light curve
that are within 3$\sigma$ of the best planet fit. We use a similar
criterion to estimate the {\it a priori} planet frequency by counting
the KOIs in the \citep{Borucki:11b} sample that have radii within
3$\sigma$ of the best fit from a planet model ($R_p =
2.227_{-0.057}^{+0.052}\,R_{\earth}$; see Table~\ref{tab:SystemParams}
below). We find that 157 among the 1235 KOIs are in this radius range
(2.06--2.38\,$R_{\earth}$), giving PF $= 157/156,\!453 = 0.0010$. This
results in a false alarm rate for \koi\ of ${\rm FAR} = 1.6 \times
10^{-5}$, which is so small that it allows us to validate the
candidate with a very high level of confidence. The planet is
designated Kepler-10\,c.

This result rests heavily on the {\it a priori} frequency of planets
from the \kepler\ Mission, derived from the assumption that all 1235
candidates reported by \cite{Borucki:11b} are indeed planets rather
than false positives.  If we were to be as pessimistic as to assume
that as many as 90\% of the small-size candidates are actually false
positives (a similar rate of false positives as is typically found in
ground-based surveys for transiting planets), and at the same time
that all of the larger-size candidates that come into the blend
frequency calculation are real planets (thereby maximizing BF and
minimizing PF), the false alarm rate would be 10 times larger than
before, or $1.6 \times 10^{-4}$. This is still a very small number,
and our conclusion regarding validation is unchanged. We note that a
rate of false positives as high as 90\% yields a planet frequency that
is strongly inconsistent not only with the expectations of
\cite{Borucki:11b} and \cite{Morton:11}, but also with the independent
results of ground based Doppler surveys as reported by
\cite{Howard:10}.

In the above calculations we have implicitly assumed similar period
distributions for planets of all sizes and for eclipsing binaries.
However, it is conceivable that the results could change if the period
distribution of planets such as Kepler-10\,c were significantly
different from the one for larger planets that go into the blend
frequency calculations, or from the one for EBs (which have a smaller
contribution to BF; see Table~\ref{tab:new_stats}).  Therefore, as a
further test we considered the impact of restricting the periods to be
within an arbitrary factor of two of the Kepler-10\,c period of 45.3
days, both in our blend frequency calculations and for the {\it a
priori} estimate of the planet frequency, PF. We find that the planet
frequencies are reduced by a factor of 4.5, and the eclipsing binary
frequency by a factor of 10.4, and as a result the false alarm rate
for \koi\ is ${\rm FAR} = 1.4 \times 10^{-5}$, which is about the same
as before. Thus, our conclusions are robust against assumptions about
the period distributions.

Finally, our false alarm rate is conservative in the sense that we
have not accounted for the flatness (coplanarity) of the Kepler-10
system. Only a small fraction of single transiting planets with
periods as long as 45 days orbiting background stars (i.e., those
acting as blends) are likely to transit, {\it a priori}, whereas a
planet of this period such as Kepler-10\,c is much more likely to
transit if it is coplanar with Kepler-10\,b. Taking this into account
would boost the planet frequency (PF) and decrease the FAR by as much
as an order of magnitude \citep[see, e.g.,][]{Beatty:10}. Coplanarity
in multiple systems is in fact supported by the large number of
multiple transiting system candidates found by \kepler\
\citep{Borucki:11b, Latham:11}, and their mutual inclinations seem to
be small \citep[1--5\arcdeg;][]{Lissauer:11b}. Therefore, we consider
our estimate of the FAR for Kepler-10\,c to be conservative.

\section{Discussion}
\label{sec:discussion}

The stellar, orbital, and planetary parameters inferred for the system
as determined by \cite{Batalha:11} are summarized in
Table~\ref{tab:SystemParams}, to which we add the transit duration.
The small formal uncertainty in the planetary radius ($\sim$2.4\%)
derives from the relatively high precision of the stellar radius,
which is based on asteroseismic constraints on the mean density of the
star.  With its radius of about 2.2\,$R_{\earth}$, Kepler-10\,c is
among the smallest exoplanets discovered to date. The mass is
undetermined as the Doppler signature has not been detected.
Nevertheless, \cite{Batalha:11} placed a constraint on it based on the
distribution of masses resulting from the Markov Chain Monte Carlo
fitting procedure they applied to the existing radial-velocity
measurements of Kepler-10.  Their conservative 3-$\sigma$ upper limit
for the mass is 20\,$M_{\earth}$.  The corresponding maximum mean
density is 10\,g~cm$^{-3}$.

Given a precise radius measurement and mass upper limit of
20\,$M_{\earth}$, some minimal constraints can be placed on the
composition of Kepler-10\,c.  Using the models of \cite{Fortney:07},
we find that an Earth-like rock-iron composition is only possible at
$\sim$ 20 $M_{\earth}$.  Lower masses would require a depletion in
iron compared to rock, or more likely an enrichment in low-density
volatiles such as water and/or H$_2$/He gas.  A 50/50 rock/water
composition yields 2.23\,$R_{\earth}$ at 7\,$M_{\earth}$.  Still lower
masses are possible with a H$_2$/He gas envelope.  Using models
presented in \cite{Lissauer:11a}, a planet with a rock/iron core and a
5\% H$_2$/He atmosphere (by mass) matches the measured radius of
Kepler-10\,c at only 3\,$M_{\earth}$.  A massive 20\,$M_{\earth}$ core
should have attained a H$_2$/He envelope, and it would appear to be
stable at Kepler-10\,c's relatively modest irradiation level, which
would lead to a planetary radius dramatically larger than
2.23\,$R_{\earth}$.  This would tend to favor a scenario where
Kepler-10\,c is more akin to GJ~1214b \citep{Charbonneau:09, Bean:10,
Nettelmann:10} and Kepler-11\,b and Kepler-11\,f, which are all below
7\,$M_{\earth}$ and enriched in volatiles.

The well measured inclinations of both Kepler-10\,b and Kepler-10\,c
allow us to put a weak constraint on the true mutual inclination
($\phi_{\rm bc}$) between the orbital planes of the two
planets. Although the relative orientation in the plane of the sky
(i.e., the mutual nodal angle) is unknown, the different impact
parameters and resulting apparent inclinations place a lower limit on
$\phi_{\rm bc}$. As discussed by \cite{Ragozzine:10}, the geometric
limits to the mutual inclination are given by $|i_b - i_c| \le
\phi_{\rm bc} \le i_b+i_c$, where $i_b = 84\fdg4^{+1.1}_{-1.6}$
\citep{Batalha:11} and $i_c = 89\fdg65^{+0.09}_{-0.12}$
(Table~\ref{tab:SystemParams}) are the usual inclinations with respect
to the line of sight.  Assuming a random orientation of the lines of
nodes (which does not account for the \emph{a priori} knowledge that
both planets are transiting), the mutual inclination is constrained to
be in the interval $5\fdg25 \le \phi_{bc} \le 174\fdg05$, with the
most likely values being at the extremes of this distribution. Making
the reasonable supposition of non-retrograde orbits, a mutual
inclination close to the lower limit of about 5\arcdeg\ is most likely
for these planets. A more detailed probabilistic argument requires
making assumptions about the number of planets in the Kepler-10
system.

This mutual inclination is on the high end of the distribution
inferred for other \kepler\ multiple candidate systems (1--5\arcdeg)
by \cite{Lissauer:11b}. If this mutual inclination is typical for
planets in this system, then it is relatively likely (depending on the
orbital period) that other planets, if present, are not
transiting. When considering the set of \kepler\ candidates in
multiple systems that have periods less than 125 days, the ratio of
periods between Kepler-10\,c and Kepler-10\,b (which is 54.1) is by
far the highest of all period ratios of neighboring pairs of \kepler\
candidates (the next highest being 23.4), and is even higher than the
period ratios between non-neighboring planets. Clearly, there is room
for multiple additional planets between Kepler-10\,b and
Kepler-10\,c. The preponderance of tightly-packed \kepler\ multiple
candidate systems suggests that additional planets may exist, and
these may be revealed in the future with more detailed transit timing
variation measurements.

Kepler-10\,c is the first \kepler\ target observed with \wspitzer\
with the aim of testing the wavelength dependence of the transit
depth. This is currently the only facility available that has the
capability of detecting such shallow transits at wavelengths that are
sufficiently separated from the \kepler\ passband to be helpful. In
this case the observations were successful, and the transit at
4.5\,\micron\ is shown to have virtually the same depth as in the
optical.  This places a very strong constraint on the color of
potential blends, which are restricted to have secondaries of similar
spectral type as the primary star.

The detailed analysis of the \kepler\ photometry with \blender\
combined with constraints from other observations eliminates the vast
majority of possible blend scenarios.  This includes most background
eclipsing binaries (leaving only a small range of possible spectral
types and relative fluxes for the secondaries), most of the scenarios
involving chance alignments with a star transited by a larger planet,
and all possible hierarchical triple configurations. The latter are
among the most difficult to detect observationally since they are
typically spatially unresolved. The key factors that have allowed
this, and made possible the validation of the planet, are the
high-precision of the \kepler\ photometry, the relatively short
ingress and egress phases (which places strong constraints on the size
ratio between the secondary and tertiary), and the near equatorial
orientation, resulting in a relatively flat transit that leaves less
freedom for the parameters of the eclipsing binaries. We expect
\blender\ to be similarly effective for other \kepler\ candidates that
show similar features in their light curves.

Kepler-10\,c along with Kepler-9\,d and Kepler-11\,g are examples of
transiting planets that have not received the usual confirmation by
dynamical means that previous discoveries have enjoyed (including
essentially all ground-based discoveries), in which either the Doppler
signature is detected unambiguously (and verified by the lack of
bisector span variations), or transit timing variations in a multiple
system are directly measured (as in Kepler-9\,b and c as well as the
five inner planets of the Kepler-11 system). Instead, the planets in
those three cases have been \emph{validated} statistically, with a
Bayesian approach to estimate the probability that the transit signals
are due to a planet rather than a false positive. This probability has
been computed by first estimating the {\it a priori} likelihood of a
false positive, and then comparing it with the {\it a priori} chance
of having observed a true planet. In the three cases mentioned above
the ratio of the false positive to planet likelihoods is small enough
that the planetary nature of the signal is established with a very
high degree of confidence. For Kepler-10\,c the false alarm rate is
$1.6 \times 10^{-5}$.

The recent work of \cite{Morton:11} has provided a means of assessing
a rough false alarm rate for \kepler\ candidates as a function of the
depth of the transit signal and the brightness of the object.  As
noted also by those authors, while these estimates are extremely
valuable for statistical studies, the validation of candidates on an
individual basis with a sufficiently high degree of confidence will
usually require a much more detailed analysis of false positives, such
as we have performed here. Masses for these objects (other than upper
limits) may of course be difficult or impractical to determine in many
cases, but it is worth keeping in mind that some of the most exciting
candidates to be discovered by \kepler\ will be in this category,
namely, Earth-size planets in the habitable zones of their parent
stars. Except for stars of late spectral type, the RV signals will
generally be very challenging to detect with the sensitivity of
current instrumentation.  Thus, statistical validation of planets is
likely to play an important role for \kepler\ in the years to come.

\acknowledgements

Funding for this Discovery mission is provided by NASA's Science
Mission Directorate. This research has made use of the facilities at
the NASA Advanced Supercomputing Division (NASA Ames Research Center),
and is based also on observations made with the Spitzer Space
Telescope which is operated by the Jet Propulsion Laboratory,
California Institute of Technology under a contract with NASA.
Support for this work was provided by NASA through an award issued by
JPL/Caltech. We thank Mukremin Kilic and Rosanne Di Stefano for
helpful discussions about white dwarfs, and the anonymous referee for
constructive comments.

{\it Facilities:} \facility{\kepler\ Mission, \wspitzer, Keck~I (HIRES),
Palomar (PHARO), WIYN}.


\clearpage



\begin{landscape}
\begin{deluxetable}{ccccccccccc}
\tabletypesize{\scriptsize}
\tablewidth{0pc}
\tablecaption{Blend frequency estimate for KOI-072.02.\label{tab:new_stats}}
\tablehead{
 & & \multicolumn{4}{c}{Blends Involving Stellar Tertiaries}
 & & \multicolumn{4}{c}{Blends Involving Planetary Tertiaries} \\ [+1.5ex]
\cline{3-6} \cline{8-11} \\ [-1.5ex]
\colhead{$K\!p$ Range} &
\colhead{$\Delta K\!p$} &
\colhead{Stellar Density\tablenotemark{a}} &
\colhead{$\rho_{\rm max}$} &
\colhead{Stars} &
\colhead{EBs} & &
\colhead{Stellar Density\tablenotemark{a}} &
\colhead{$\rho_{\rm max}$} &
\colhead{Stars} &
\colhead{Transiting Planets}
\\
\colhead{(mag)} &
\colhead{(mag)} &
\colhead{(per sq.\ deg)} &
\colhead{(\arcsec)} &
\colhead{($\times 10^{-6}$)} &
\colhead{$f_{\rm EB} = 0.78$\%} & &
\colhead{(per sq.\ deg)} &
\colhead{(\arcsec)} &
\colhead{($\times 10^{-6}$)} &
\colhead{0.42--1.84\,$R_{\rm Jup}$, $f_{\rm Plan}=0.17$\%}
\\
\colhead{} &
\colhead{} &
\colhead{} &
\colhead{} &
\colhead{} &
\colhead{($\times 10^{-6}$)} & &
\colhead{} &
\colhead{} &
\colhead{} &
\colhead{($\times 10^{-6}$)} 
\\
\colhead{(1)} &
\colhead{(2)} &
\colhead{(3)} &
\colhead{(4)} &
\colhead{(5)} &
\colhead{(6)} & &
\colhead{(7)} &
\colhead{(8)} &
\colhead{(9)} &
\colhead{(10)}
}
\startdata
11.0--11.5  &  0.5 &\nodata &\nodata  & \nodata & \nodata && \nodata&\nodata & \nodata& \nodata \\
11.5--12.0  &  1.0 &\nodata &\nodata  & \nodata & \nodata && \nodata&\nodata & \nodata& \nodata \\
12.0--13.0  &  1.5 &\nodata &\nodata  & \nodata & \nodata && \nodata&\nodata & \nodata& \nodata \\
12.5--13.0  &  2.0 &\nodata &\nodata  & \nodata & \nodata && \nodata&\nodata & \nodata& \nodata \\
13.0--13.5  &  2.5 &\nodata &\nodata  & \nodata & \nodata &&  139   &  0.12  &  0.485 & 0.0008  \\
13.5--14.0  &  3.0 & 32     & 0.15    & 0.175   & 0.0014  &&  197   &  0.15  &  1.074 & 0.0018  \\
14.0--14.5  &  3.5 & 44     & 0.18    & 0.346   & 0.0027  &&  278   &  0.18  &  2.183 & 0.0037  \\
14.5--15.0  &  4.0 &\nodata &\nodata  & \nodata & \nodata &&  351   &  0.20  &  3.403 & 0.0058  \\
15.0--15.5  &  4.5 &\nodata &\nodata  & \nodata & \nodata && \nodata&\nodata & \nodata& \nodata \\
15.5--16.0  &  5.0 &\nodata &\nodata  & \nodata & \nodata && \nodata&\nodata & \nodata& \nodata \\
16.0--16.5  &  5.5 &\nodata &\nodata  & \nodata & \nodata && \nodata&\nodata & \nodata& \nodata \\
16.5--17.0  &  6.0 &\nodata &\nodata  & \nodata & \nodata && \nodata&\nodata & \nodata& \nodata \\
17.0--17.5  &  6.5 &\nodata &\nodata  & \nodata & \nodata && \nodata&\nodata & \nodata& \nodata \\
17.5--18.0  &  7.0 &\nodata &\nodata  & \nodata & \nodata && \nodata&\nodata & \nodata& \nodata \\
18.0--18.5  &  7.5 &\nodata &\nodata  & \nodata & \nodata && \nodata&\nodata & \nodata& \nodata \\
18.5--19.0  &  8.0 &\nodata &\nodata  & \nodata & \nodata && \nodata&\nodata & \nodata& \nodata \\
\noalign{\vskip 6pt}
\multicolumn{2}{c}{Totals} & 76 &\nodata &  0.521 & {\bf 0.0041} && 965  &\nodata & 7.145 & {\bf 0.0121}   \\
\noalign{\vskip 4pt}
\hline
\noalign{\vskip 4pt}
\multicolumn{11}{c}{Blend frequency (BF) = $(0.0041 + 0.0121)\times 10^{-6}= 1.62 \times 10^{-8}$} \\ [-1.5ex]
\enddata

\tablenotetext{a}{The number densities in columns 3 and 7 differ
because of the different secondary mass ranges permitted by \blender\
for the two kinds of blend scenarios, as shown in
Figure~\ref{fig:backstar} and Figure~\ref{fig:back_plan}.}
\tablecomments{Magnitude bins with no entries correspond to brightness ranges in
which all blends are ruled out by a combination of \blender\ and other constraints.}

\end{deluxetable}
\clearpage
\end{landscape}

\clearpage

\begin{deluxetable}{lcc}
\tabletypesize{\scriptsize}
\tablewidth{0pc}
\tablecaption{Star and planet parameters for Kepler-10\,c.\label{tab:SystemParams}}
\tablehead{
\colhead{Parameter}  &
\colhead{Value}      &
\colhead{Notes}
}
\startdata

\sidehead{\em Spectroscopically determined stellar parameters}
Effective temperature, $T_{\rm eff}$ (K)         & $5627 \pm 44$\phn\phn  & A     \\
Surface gravity, $\log g$ (cgs)                  & $4.35 \pm 0.06$        & A     \\
Metallicity, [Fe/H]                              & $-0.15 \pm 0.04$\phs   & A     \\
Projected rotation, $v \sin i$ (\kms)            & $0.5 \pm 0.5$          & A     \\

\sidehead{\em Inferred host star properties}
Mass, $M_{\star}$ ($M_{\sun}$)                   & $0.895 \pm 0.060$      & B     \\
Radius, $R_{\star}$ ($R_{\sun}$)                 & $1.056 \pm 0.021$      & B     \\
Surface gravity, $\log g_{\star}$ (cgs)          & $4.341 \pm 0.012$      & B     \\
Luminosity, $L_{\star}$ ($L_{\sun}$)             & $1.004 \pm 0.059$      & B     \\
Absolute $V$ magnitude, $M_V$ (mag)              & $4.746 \pm 0.063$      & B     \\
Age (Gyr)                                        & $11.9 \pm 4.5$\phn     & B     \\
Distance (pc)                                    & $173 \pm 27$\phn       & B     \\

\sidehead{\em Transit and orbital parameters}
Orbital period, $P$ (days)                       & $42.29485_{-0.00076}^{+0.00065}$         & C     \\
Mid-transit time, $T_c$ (HJD)                     & $2,\!454,\!971.6761_{-0.0023}^{+0.0020}$ & C     \\
Scaled semimajor axis, $a/R_{\star}$             & $49.1_{-1.3}^{+1.2}$                     & C     \\
Scaled planet radius, $R_{\rm p}/R_{\star}$      & $0.01938_{-0.00024}^{+0.00020}$          & C     \\
Impact parameter, $b$                            & $0.299_{-0.073}^{+0.089}$                & C     \\
Orbital inclination, $i$ (deg)                   & $89.65_{-0.12}^{+0.09}$                  & C     \\
Transit duration, $\Delta$ (hours)               & $6.863_{-0.068}^{+0.065}$                & C     \\

\sidehead{\em Parameters for Kepler-10\,c}
Radius, $R_{\rm p}$ ($R_{\earth}$)               & $2.227_{-0.057}^{+0.052}$                & B,C     \\
Mass, $M_{\rm p}$ ($M_{\earth}$)                 & $< 20$                                   & D       \\
Mean density, $\rho_{\rm p}$ (g~cm$^{-3}$)       & $< 10$                                   & D       \\
Orbital semimajor axis, $a$ (AU)                 & $0.2407_{-0.0053}^{+0.0044}$             & E       \\
Equilibrium temperature, $T_{\rm eq}$ (K)        & 485                                      & F \\ [-1.5ex]

\enddata
\tablecomments{
In most cases these parameters are taken from \cite{Batalha:11}.
A: Based on an analysis by D.\ Fischer of the Keck/HIRES template
spectrum using SME \citep[see][]{Valenti:96, Batalha:11};
B: Based on the asteroseismology analysis and stellar models;
C: Based on an analysis of the photometry;
D: Upper limit corresponding to three times the 68.3\% credible interval 
from the MCMC mass distribution;
E: Based on Newton's revised version of Kepler's Third Law and the 
results from D;
F: Calculated assuming a Bond albedo of 0.1 and complete redistribution 
of heat for re-radiation.
}

\end{deluxetable}


\begin{thebibliography}



\bibitem[Batalha et al.(2010)]{Batalha:10}
 Batalha, N.\ M.\ et al.\ 2010, \apj, 713, L103

\bibitem[Batalha et al.(2011)]{Batalha:11} 
 Batalha, N.\ M.\ et al.\ 2011, \apj, 729, 27

\bibitem[Bean et al.(2010)]{Bean:10}
 Bean, J.\ L., Kempton, E.\ M.-R., \& Homeier, D. 2010, \nat, 468, 669

\bibitem[Beatty \& Seager(2010)]{Beatty:10}
 Beatty, T.\ G., \& Seager, S. 2010, \apj, 712, 1433

\bibitem[Beerer et al.(2011)]{beerer11}
 Beerer, I.\ M., et al.\ 2011, \apj, 727, 23 


\bibitem[Borucki et al.(2011a)]{Borucki:11a}
 Borucki, W.\ J.\ et al.\ 2011a, \apj, 728, 117

\bibitem[Borucki et al.(2011b)]{Borucki:11b} 
 Borucki, W.\ J.\ et al.\ 2011b, arXiv:1102.0541 

\bibitem[Brown et al.(2011)]{Brown:11}
 Brown, T.\ M., Latham, D.\ W., Everett, M.\ E., \& Esquerdo, G.\ A.,
 \aj, submitted (arXiv:1102.0342)




\bibitem[Carter et al.(2011)]{Carter:11}
 Carter, J.\ A., Rappaport, S., \& Fabrycky, D. 2011, \apj, 728, 139


\bibitem[Charbonneau et al.(2009)]{Charbonneau:09}
 Charbonneau, D.\ et al.\ 2009, \nat, 462, 891

\bibitem[Charbonneau et al.(2005)]{charbonneau05}
 Charbonneau, D.\ et al. 2005, \apj, 626, 523

\bibitem[Claret(2000)]{Claret:00}
 Claret, A. 2000, \aap, 363, 1081

\bibitem[D{\'e}sert et al.(2009)]{desert09}
 D{\'e}sert, J.-M., Lecavelier des Etangs, A., H{\'e}brard, G., Sing,
 D.~K., Ehrenreich, D., Ferlet, R., \& Vidal-Madjar, A.\ 2009, \apj,
 699, 478

\bibitem[D{\'e}sert et al.(2011a)]{desert11a}
 D{\'e}sert, J.-M.\ et al.\ 2011, \aap, 526, A12 

\bibitem[D{\'e}sert et al.(2011b)]{desert11b}
 D{\'e}sert, J.-M.\ et al.\ 2011, \apj, submitted (arXiv:1102.0555)

\bibitem[Deming et al.(2011)]{deming11}
 Deming, D.\ et al.\ 2011, \apj, 726, 95 

\bibitem[Eastman et al.(2010)]{eastman10}
 Eastman, J., Siverd, R., \& Gaudi, B.\ S.\ 2010, \pasp, 122, 935 



\bibitem[Fazio et al.(2004)]{fazio04} Fazio, G.\ G.\ et al. 2004, \apjs, 154, 10


\bibitem[Fortney et al.(2007)]{Fortney:07}
 Fortney, J.\ J., Marley, M.\ S., \& Barnes, J.\ W. 2007, \apj, 659, 1661







\bibitem[Holman et al.(2010)]{Holman:10}
 Holman, M.\ J.\ et al. 2010, Science, 330, 51


\bibitem[Howard et al.(2010)]{Howard:10} 
 Howard, A.\ W.\ et al.\ 2010, Science, 330, 653 

\bibitem[Howard et al.(2011)]{Howard:11} 
 Howard, A.\ W.\ et al.\ 2011, in preparation

\bibitem[Howell et al.(2011)]{Howell:11}
 Howell, S.\ B.\ et al.\ 2011, \aj, submitted





\bibitem[Knutson et al.(2008)]{knutson08}
 Knutson, H.\ A., Charbonneau, D., Allen, L.\ E., Burrows, A., \&
 Megeath, S.\ T.\ 2008, \apj, 673, 526



\bibitem[Latham et al.(2011)]{Latham:11}
 Latham, D.\ W.\ et al.\ 2011, \apjl, in press (arXiv:1103.3896)

\bibitem[L\'eger et al.(2009)]{Leger:09}
 L\'eger, A.\ et al.\ 2009, \aap, 506, 287

\bibitem[Lissauer et al.(2011a)]{Lissauer:11a}
 Lissauer, J.\ J.\ et al.\ 2011a, \nat, 470, 53

\bibitem[Lissauer et al.(2011b)]{Lissauer:11b}
 Lissauer, J.\ J.\ et al.\ 2011b, \apj, submitted (arXiv:1102.0543)


\bibitem[Mandel \& Agol(2002)]{mandel02} 
 Mandel, K., \& Agol, E.\ 2002, \apj, 580, L171 



\bibitem[Markwardt(2009)]{markwardt09}
 Markwardt, C.\ B.\ 2009, Astronomical Society of the Pacific
 Conference Series, 411, 251


\bibitem[Mazeh(2008)]{Mazeh:08} 
Mazeh, T. 2008, in Tidal Effects in Stars, Planets and Disks, EAS
 Publications Series, eds.\ M.-J.\ Goupil \& J.-P.\ Zahn (EDP
 Sciences), Vol.\ 29, p. 1


\bibitem[Moorhead et al.(2011)]{Moorhead:11}
 Moorhead, A.\ V.\ et al.\ 2011, arXiv:1102.0547


\bibitem[Morton \& Johnson(2011)]{Morton:11} 
Morton, T.\ D., \& Johnson, J.\ A. 2011, arXiv:1101.5630 


\bibitem[Nettelmann et al.(2010)]{Nettelmann:10}
 Nettelmann, N., Fortney, J.\ J., Kramm, U., \& Redmer, R. 2010, \apj,
 in press (arXiv:1010.0277)


\bibitem[Panei et al.(2000)]{Panei:00}
 Panei, J.\ A., Althaus, L.\ G., \& Benvenuto, O.\ G. 2000, \aap, 353, 977

\bibitem[Panei et al.(2007)]{Panei:07}
 Panei, J.\ A., Althaus, L.\ G., Chen, X., \& Han, Z. 2007, \mnras, 382, 779

\bibitem[Pont et al.(2006)]{pont06}
 Pont, F., Zucker, S., \& Queloz, D.\ 2006, \mnras, 373, 231 


\bibitem[Press et al.(1992)]{Press:92}
Press, W.\ H., Teukolsky, S.\ A., Vetterling, W.\ T., \& Flannery, B.\
P. 1992, {\it Numerical Recipes}, (2nd.\ Ed.; Cambridge: Cambridge
Univ.\ Press), 650




\bibitem[Ragozzine \& Holman(2010)]{Ragozzine:10}
 Ragozzine, D., \& Holman, M.\ J. 2010, \apj, submitted (arXiv:1006.3727)

\bibitem[Robin et al.(2003)]{Robin:03}
 Robin, A.\ C., Reyl\'e, C., Derri\'ere, S., \& Picaud, S. 2003, \aap,
 409, 523

\bibitem[Rowe et al.(2010)]{Rowe:10}
 Rowe, J.\ F.\ et al.\ 2010, \apj, 713, L150

\bibitem[Sing(2010)]{Sing:10}
 Sing, D.\ K. 2010, \aap, 510, 21

\bibitem[Slawson et al.(2011)]{Slawson:11}
 Slawson, R.\ W.\ et al.\ 2011, \aj, submitted (arXiv:1103.1659)

\bibitem[Snellen et al.(2009)]{Snellen:09}
 Snellen, I.\ A.\ G.\ et al.\ 2009, \aap, 497, 545


\bibitem[Torres et al.(2004)]{Torres:04}
 Torres, G., Konacki, M., Sasselov, D.\ D., \& Jha, S. 2004, \apj,
614, 979


\bibitem[Torres et al.(2011)]{Torres:11} 
 Torres, G.\ et al.\ 2011, \apj, 727, 24 


\bibitem[Valenti \& Piskunov(1996)]{Valenti:96}
 Valenti, J.\ A., \& Piskunov, N. 1996, \aap, 118, 595


\bibitem[Werner et al.(2004)]{werner04} Werner, M.\ W., et al.\ 2004, \apjs, 154, 1

\bibitem[Winn(2010)]{Winn:10}
 Winn, J.\ N.\ 2010, in Exoplanets, ed.\ S.\ Seager (Tucson: Univ.\ of
 Arizona Press), p.\ 55


\end{thebibliography}
\end{document}